\documentclass[11pt]{article}
\usepackage[letterpaper,margin=1in]{geometry}
\usepackage{amsmath,amssymb,bm}
\usepackage{graphicx}
\usepackage{subeqnarray}
\usepackage{titling}
\predate{}\postdate{}
\usepackage{booktabs}

\numberwithin{equation}{section}
\usepackage[numbers,sort&compress]{natbib}
\usepackage[colorlinks=true,citecolor=blue,linkcolor=blue,urlcolor=blue]{hyperref}

\newcommand{\Rey}{\mathit{Re}}

\newcommand{\bnabla}{\boldsymbol{\nabla}}
\newcommand{\bcdot}{\boldsymbol{\cdot}}
\newenvironment{keywords}{}{}
\newenvironment{bmhead}[1][]{\par\bigskip\noindent\textbf{#1}\hspace{0.5em}}{\par}

\title{Is No-Slip Necessary for Vorticity Generation and Shedding over a Circular Cylinder?}
\author{Kourosh Jafari Ghalejooghi \qquad Haithem E. Taha\thanks{Corresponding author: \texttt{hetaha@uci.edu}}\\[8pt]
\normalsize Department of Mechanical and Aerospace Engineering,\\
\normalsize University of California, Irvine, CA 92697, USA}
\date{}

\begin{document}
\maketitle
\vspace{-2.5em}

\begin{abstract}
It is traditionally believed that the no-slip boundary condition is necessary for vorticity generation, as might be implied by the Lighthill vorticity generation mechanism. However, the investigations of \citet{morton1984generation} and \citet{terrington2020generation} assert that the vorticity-generation mechanism is independent of the no-slip boundary condition. To investigate this hypothesis, we simulate viscous flow over a circular cylinder in the laminar periodic von K\'{a}rm\'{a}n vortex-shedding regime at $\Rey=170$, without enforcing the no-slip condition at the wall. Instead, we close the wall tangential velocity and its normal derivative using one-sided finite-difference approximations. Accordingly, the wall tangential velocity is dynamically determined from the evolving interior solution rather than prescribed. The resulting flows exhibit qualitatively similar patterns of periodic vortex shedding across all three wall treatments (no-slip, second- and third-order one-sided approximations), although the mean drag coefficient, r.m.s.\ lift coefficient, and Strouhal number differ from those of the standard no-slip simulation. More significantly, the velocity distributions just outside the boundary layer closely match those of the standard no-slip simulation, regardless of the approximation order. In particular, the circulation evaluated along a contour at the edge of the boundary layer, which corresponds to the total vorticity contained within the layer in the no-slip case, is found to be remarkably insensitive to the wall treatment. These results suggest that the no-slip condition is not necessary for vorticity generation or periodic shedding. However, it may still be required for accurate quantitative prediction of the integral flow quantities and shedding dynamics.
\end{abstract}


\section{Introduction}
\label{sec:intro}
The vorticity generation mechanism is one of the most fundamental concepts in fluid mechanics. Yet, it has been a subject of debate over the last several decades. There are two competing schools of thought: one identifies viscosity and the no-slip boundary condition as the main enablers of vorticity generation (\citet{Wu_Vorticity_Generation,Viscous_Freq_Resp,De3z_Seeny_Vorticity_Generation,De3z_Seeny_Kutta}), while the other argues that the mechanism is inviscid and independent of the no-slip boundary condition (\citet{morton1984generation,brons2014vorticity,terrington2020generation}). 

In fact, the traditional view presented in classical textbooks (e.g., \citet{lighthill1963introduction,batchelor1967introduction}) made it clear that viscosity (represented by the diffusion term) cannot \textit{create fresh} vorticity that did not previously exist. For example, in Sec.~1.5, `Solid Boundaries as Sources of Vorticity' (p. 54), \citet{lighthill1963introduction} acknowledged the central fact on which the viscous school relies: ``\textit{The ... conclusion ... that a particle of fluid with zero vorticity will continue to have zero vorticity, is false in a viscous fluid, since diffusion of vorticity from nearby particles can occur}.'' However, he immediately clarified in the very next statement: ``\textit{But diffusion cannot create vorticity out of nothing, so that in external aerodynamics one may reasonably ask: when a uniform stream flows past an obstacle, how is vorticity imparted to the fluid, all of which lacks it initially?}'' He then answered: 
``\textit{The answer is that the solid boundary is a `distributed source' of vorticity}.''

Lighthill then asked ``\textit{how does a body moving through otherwise undisturbed fluid determine the development of the flow around it?}'' To answer this question, he considered an impulsive start, arguing that, at $t=0$, there is no effect of viscous diffusion and the initial flow is irrotational, governed only by the inviscid Euler dynamics and the no-penetration boundary condition. He then asserted that ``\textit{the resulting tangential velocity at the surface may not satisfy the no-slip condition. In this case, we deduce that exactly enough tangential vorticity must have been created at the surface ... so that the velocity field of that vorticity ... combines with that previously determined to give zero slip}.'' That is, the total amount of generated vorticity is exactly that required to bridge the difference between the inviscid tangential velocity $U_e$ and the surface velocity $U_s$: 
\begin{equation}\label{eq:gamma}
\gamma(x) = \int_0^\delta \omega(x,y) dy = U_e(x) - U_s(x),
\end{equation}
where $x$, $y$ are the tangential and normal coordinates, respectively, and $\delta$ is the boundary layer thickness. Thus, the integrated amount of vorticity generated is determined kinematically by the velocity difference $\gamma=U_e-U_s$, with no explicit dependence on viscosity. This is commonly referred to as the \textit{Lighthill vorticity generation mechanism}.

\citet{batchelor1967introduction} gave a similar exposition to Lighthill's. In Sec.~5.4, `The Source of Vorticity in Motions Generated From Rest' (p. 277), Batchelor asserted that the initial flow following an impulsive start is irrotational, which ``\textit{is determined completely by the condition of zero flux of mass across ... the solid boundary [i.e., no penetration], and this unique irrotational motion almost inevitably has a non-zero tangential component of relative velocity of the fluid at the solid boundary ... Thus, the motion that would be generated from rest in the absence of diffusion of vorticity across the boundary of the fluid is accompanied by a non-zero tangential relative velocity at the boundary}.'' He asserted that ``\textit{the vorticity in this flow is infinite at the boundary.}'' That is, the jump from $U_s$ to $U_e$ occurs across zero thickness, resulting in a sheet of infinite vorticity. Batchelor further clarified: ``\textit{This sheet of infinite vorticity at the boundary is the source from which, once viscosity is allowed to act, vorticity diffuses into the interior of the fluid.}''

As described above, both Lighthill and Batchelor asserted that viscous diffusion plays no role in vorticity generation itself, and that the total amount of vorticity is independent of the value of viscosity; rather, it is dictated by the surface velocity and the inviscid dynamics. From this perspective, they support the inviscid school of \citet{morton1984generation,brons2014vorticity,terrington2020generation}. However, it is clear that the Lighthill vorticity generation mechanism invokes the no-slip boundary condition, specifically through the jump $\gamma$ between the inviscid tangential velocity distribution $U_e$ and the surface velocity $U_s$, with the latter imposed by \textit{stickiness} (i.e., the no-slip condition). Hence, from this perspective, the Lighthill mechanism is not purely inviscid: an arbitrarily small but nonzero viscosity is still needed to enforce stickiness. In other words, the Lighthill mechanism is independent of the \textit{magnitude} of viscosity, but not of its \textit{presence}, in contrast to the claims of \citet{morton1984generation,brons2014vorticity,terrington2020generation}. 

Interestingly, the no-slip condition itself was the subject of debate throughout much of the nineteenth century (see \citet{goldstein1938modern,neto2005boundary}, and the references therein). Several prominent scholars rejected, or at least questioned, the no-slip assumption. \citet{navier1823memoire} proposed a partial-slip law, in which the tangential velocity at the wall is proportional to the wall shear stress. \citet{stokes1845theories} was initially inclined towards the no-slip condition, but, when computing the discharge through pipes and canals, found that the resulting formulae ``\textit{did not at all agree with experiment}'', and concluded that the conditions at the surface of a solid ``\textit{except perhaps in case of very small motions, are unknown}''. As described by \citet{goldstein1938modern}, ``\textit{Gradually ... the hypothesis finally adopted by Stokes, that there is no slip, prevailed}'', based on the reasoning that finite slip was ``\textit{exceedingly improbable a priori}'' and on the agreement between theoretical predictions based on the no-slip condition and experimental observations (\citet{stokes1851effect}). However, as also emphasised by \citet{goldstein1938modern}, ``\textit{All these experiments ... relate almost entirely to non-turbulent flow.}'' Thus, the no-slip condition ultimately became the standard wall closure for viscous flows (\citet{goldstein1938modern}). Importantly, however, it was not established as a consequence of first principles, but rather adopted as a closure whose macroscopic predictions, mainly in the laminar regime, agree with experiment (\citet{neto2005boundary}).

The goal of this paper is to perform a systematic study of the role of the no-slip condition in vorticity generation. In particular, we ask whether the no-slip condition is necessary for vorticity generation. To answer this question, we adopt the \textit{method of difference} by \citet{mill1843system}, a classical method of causal inference in logic and the philosophy of science. We simply eliminate the no-slip condition while keeping the remaining dynamics unchanged, and investigate whether vorticity is still generated. 

We choose the unsteady laminar flow over a circular cylinder at $\Rey = 170$ as a case study, for which periodic vortex shedding is expected in the physical flow. We simulate the two-dimensional incompressible Navier--Stokes equations both with and without the no-slip boundary condition and compare the resulting flow fields. The results of this study should provide insight into whether the no-slip boundary condition is an essential mechanism for vorticity generation and periodic shedding in flow past a circular cylinder, and whether resolving the boundary layer is necessary to reproduce the qualitative features of the flow field.

\section{Problem Formulation}
\label{sec:formulation}

The simulations are performed by numerically solving the two-dimensional incompressible Navier--Stokes equations in polar coordinates. The momentum equations in the $r$ and $\theta$ directions are, respectively:
\begin{equation}
\frac{\partial u_r}{\partial t} 
+ u_r \frac{\partial u_r}{\partial r} 
+ \frac{u_\theta}{r}\frac{\partial u_r}{\partial \theta} 
- \frac{u_\theta^2}{r} 
= -\frac{1}{\rho}\frac{\partial p}{\partial r} 
+ \nu \left( \nabla^2 u_r - \frac{u_r}{r^2} 
- \frac{2}{r^2}\frac{\partial u_\theta}{\partial \theta} \right),
\end{equation}
\begin{equation}
\frac{\partial u_\theta}{\partial t} 
+ u_r \frac{\partial u_\theta}{\partial r} 
+ \frac{u_\theta}{r}\frac{\partial u_\theta}{\partial \theta} 
+ \frac{u_r u_\theta}{r} 
= -\frac{1}{\rho r}\frac{\partial p}{\partial \theta} 
+ \nu \left( \nabla^2 u_\theta + \frac{2}{r^2}\frac{\partial u_r}{\partial \theta} 
- \frac{u_\theta}{r^2} \right),
\end{equation}
where $p$ is the pressure, $\rho$ is the fluid density, $\nu$ is the kinematic viscosity, and $u_r$ and $u_\theta$ are the velocity components in the $r$ and $\theta$ directions, respectively. These equations are solved subject to the incompressibility constraint:
\begin{equation}
\frac{1}{r}\frac{\partial (r u_r)}{\partial r} 
+ \frac{1}{r}\frac{\partial u_\theta}{\partial \theta} = 0.
\end{equation}

The velocity and pressure fields are decoupled using the fractional-step projection method proposed by \citet{chorin1968numerical}, with first-order Euler time integration. Although the reader is assumed to be familiar with this method, we present the governing equations in vector form to highlight a key observation that further motivates the present investigation into the 
role of the no-slip boundary condition.

In the predictor step of Chorin's projection method, an intermediate velocity field $\boldsymbol{u}^*$ is computed by advancing the momentum equation in time without the 
pressure gradient term:
\begin{equation}
\frac{\boldsymbol{u}^* - \boldsymbol{u}^n}{\Delta t} 
= - (\boldsymbol{u}^n \bcdot \bnabla)\boldsymbol{u}^n 
+ \nu \bnabla^2 \boldsymbol{u}^n \equiv  \boldsymbol{a}^{\mathrm{free}},
\end{equation}
where $\boldsymbol{u}^n$ is the velocity field at time step $n$, and $\boldsymbol{a}^{\mathrm{free}}$ is the \emph{free acceleration} in the language of Gauss's principle of least constraint (\citet{taha2023minimization}): the acceleration that would take place in the absence of the continuity constraint. The pressure gradient itself enforces the incompressibility constraint. 

In the second step, the pressure field is obtained by solving the Pressure 
Poisson Equation (PPE):
\begin{equation}
\nabla^2 p^{n+1} = \frac{\rho}{\Delta t} \bnabla\bcdot\boldsymbol{u}^*.
\end{equation}
Finally, in the corrector step, the intermediate velocity field is projected onto the space of divergence-free fields using the computed pressure gradient:
\begin{equation}
\boldsymbol{u}^{n+1} = \boldsymbol{u}^* 
- \frac{\Delta t}{\rho} \bnabla p^{n+1}.
\end{equation}

In an unsteady simulation, the current velocity field $\boldsymbol{u}$ is known, while the main unknowns are its evolution $\partial\boldsymbol{u}/\partial t$ and the pressure field. It is well-established that the appropriate boundary condition for pressure at a solid surface is a Neumann boundary condition derived from the normal momentum equation, as discussed in detail by \citet{gresho1987pressure}. Hence, no tangential velocity boundary condition is needed to solve the PPE, since its source term and boundary condition depend only on the current velocity field $\boldsymbol{u}$, which is already known at a given time step. A tangential boundary condition is needed only when advancing the tangential momentum equation, specifically in evaluating the viscous terms at the wall, which involve quantities such as $\partial {u_\theta}/\partial r$. 

In the case where we do not enforce the no-slip boundary condition, this radial derivative is obtained directly from the interior solution using a one-sided approximation. Consequently, both the tangential velocity at the wall and its radial derivative are allowed to evolve naturally according to the flow dynamics, without being externally prescribed. Whether the resulting problem remains well-posed and physically meaningful is not immediately clear. 

One might argue that computing a one-sided derivative of $\partial {u_\theta}/\partial r$, which is effectively an extrapolation from the interior solution, amounts to imposing \textit{some} tangential boundary closure. While this may be true, such a closure imposes no stickiness whatsoever. As will be shown below in the results section, the tangential velocity at the wall evolves freely, without any tendency to adhere to the surface. This is precisely the behaviour under investigation.

\begin{figure}
    \centering
    \includegraphics[width=0.8\textwidth]{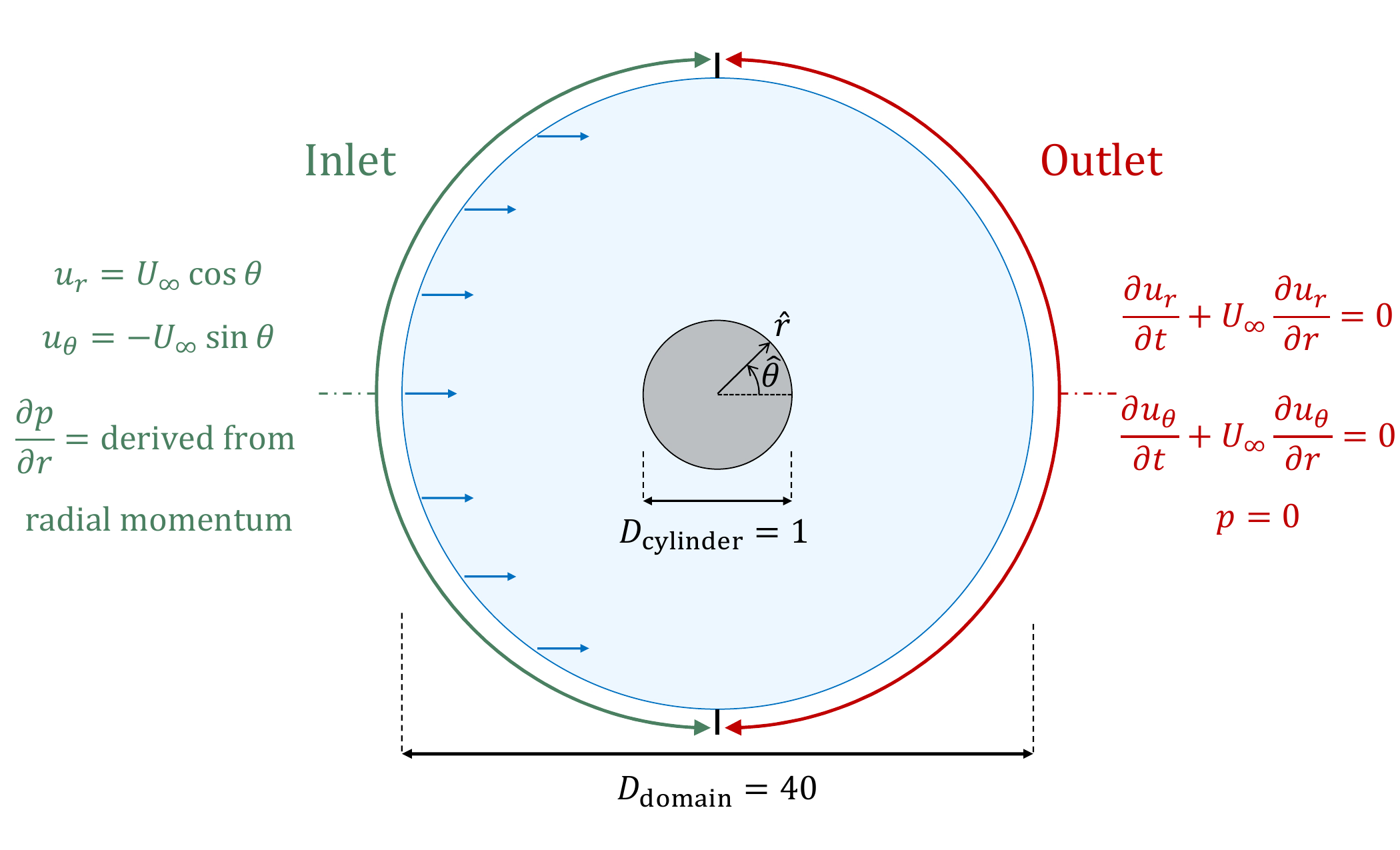}
    \caption{Schematic of the computational domain and boundary conditions for the simulation of flow over a circular cylinder.}
    \label{fig:domain_schematic}
\end{figure}

The simulations are performed at $\Rey = 170$, which, according to \citet{zdravkovich1997flow}, falls within the laminar vortex shedding regime (L3, $48 < \Rey < 180$), just below the onset of the transition-in-wake (TrW) regime in which three-dimensional instabilities lead to the formation of streamwise vortex structures.

\section{Numerical Methodology}
\label{sec:method}
All simulations are performed using an in-house finite-volume solver. Both the convective and diffusive terms are discretised using second-order central differencing, and time integration is carried out with the first-order explicit Euler scheme. The time step is set to the largest value satisfying both the advective (CFL) and diffusive stability limits, evaluated using the radial width of the smallest cell at the cylinder wall. At $\Rey = 170$, the advective constraint is binding in all cases. A CFL number of 0.4 is used for the no-slip simulation and for the slip simulation with second-order wall extrapolation (i.e., the one-sided closure introduced in §\ref{sec:formulation}). The third-order extrapolation case requires a more conservative value of 0.3 to maintain stability, owing to the larger alternating-sign coefficients of the four-point Lagrange stencil near the wall.

The pressure Poisson equation (PPE) is solved using the successive over-relaxation (SOR) method (\citet{young1954iterative}), with a fixed relaxation parameter of $1.6$, chosen empirically to balance convergence rate and stability. The PPE is iterated until the r.m.s.\ residual falls below $10^{-4}$. The pressure boundary conditions at the cylinder wall and at the inlet are the consistent Neumann conditions derived by applying the radial momentum equation at the respective boundaries, following \citet{gresho1987pressure}.

Following \citet{rajani2009numerical}, we resolve the flow in a circular domain of diameter $40D$, with the cylinder at its centre, as shown in the schematic in figure~\ref{fig:domain_schematic}.
An O-grid with non-uniform radial spacing is employed, with cells clustered near the cylinder wall. A mesh-independence study is conducted by successively refining the grid and monitoring the mean drag coefficient $\bar{C}_D$ until its variation between successive refinements falls below $6\%$. The final mesh comprises $650$ cells in the radial direction and $450$ cells in the azimuthal direction, with the radial thickness of the first cell at the wall equal to $1\times10^{-2}\,D$, where $D$ is the cylinder diameter. Further details of the mesh and the convergence study are provided in the supplementary materials.

In the standard simulation, the no-slip condition is enforced at the cylinder wall by setting both velocity components to zero. In the slip simulations, however, only the no-penetration condition is retained (i.e.\ the radial component of velocity is set to zero), while the tangential component is left unconstrained and closed via extrapolation from the interior. Consequently, the wall tangential velocity is not prescribed but is instead dynamically determined by the evolving flow field in the vicinity of the wall.

The extrapolation is performed by means of a Taylor-series expansion about $r = R_{\mathrm{cyl}}$, using the velocity values at the adjacent cell centres. The order of accuracy is controlled by the number of cells included in the stencil. The derivation of the second-order scheme is presented below, assuming a uniform radial grid; the third-order scheme follows analogously, extending the stencil to four cell centres. More information on the general case of the derivation for a non-uniform grid is provided in the supplementary materials.

The tangential velocity $u_{\theta,w}$ at the cylinder wall ($r = R_{\mathrm{cyl}}$) and its radial gradient ($u_{\theta,w}'$) are obtained by expanding $u_{\theta}$ in a Taylor series about $r = R_{\mathrm{cyl}}$ in terms of its value at the three nearest cell centres. For a uniform radial grid of cell width $\Delta r = 2h$, the cell centres lie at distances $d_{k} = (2k-1)h$ from the wall, giving 
\begin{equation}
  u_{\theta,k} = u_{\theta,w}
    + (2k-1)h\,u_{\theta,w}'
    + \frac{(2k-1)^2 h^{2}}{2}\,u_{\theta,w}''
    + O(h^{3}), \slabel{eq:Taylor2_a}
\end{equation}
where the subscript $k$ denotes the value at the $k$-th cell centre from the wall ($k = 1, 2, 3$). This constitutes a $3\times3$ linear system in the three unknowns $u_{\theta,w}$, $u_{\theta,w}'$, and $u_{\theta,w}''$. Solving for $u_{\theta,w}$ and $u_{\theta,w}'$ yields
\begin{equation}
  u_{\theta}\big|_{r=R_{\mathrm{cyl}}}
  = \frac{1}{8}\!\left(
      15\,u_{\theta,1}
    - 10\,u_{\theta,2}
    +  3\,u_{\theta,3}
    \right)
  + O(h^{3}),
  \label{eq:wall_utheta_2nd}
\end{equation}
\begin{equation}
  \left.\frac{\partial u_{\theta}}{\partial r}\right|_{r=R_{\mathrm{cyl}}}
  = \frac{1}{2h}\!\left(
    - 2\,u_{\theta,1}
    + 3\,u_{\theta,2}
    -   u_{\theta,3}
    \right)
  + O(h^{2}).
  \label{eq:wall_dutheta_2nd}
\end{equation}
The radial gradient may equivalently be obtained using a one-sided, second-order finite-difference approximation constructed on the same cell-centred stencil; this yields an expression identical to the one derived above.

We emphasise that this closure does not fix the tangential wall velocity or its gradients \textit{a priori}. Rather, they evolve freely under the system dynamics. In particular, the wall tangential velocity (\ref{eq:wall_utheta_2nd}) and its radial gradient (\ref{eq:wall_dutheta_2nd}) are determined entirely by the evolving interior solution at every time step. The approach is distinct from partial-slip boundary conditions, such as the Navier or Maxwell slip models, in which a linear relationship is prescribed between the wall tangential velocity and the wall shear stress through a constant coefficient known as the slip length. No such prescribed relationship exists between (\ref{eq:wall_utheta_2nd}) and (\ref{eq:wall_dutheta_2nd}). Thus, no kinematic stickiness is imposed at the wall: the tangential fluid velocity is not constrained to follow the wall velocity, but is allowed to evolve freely according to the interior flow dynamics.

\section{Results and Discussion}
\label{sec:resultsanddiscussion}

We first validate our standard no-slip simulation against the literature by comparing the results of the mean drag coefficient $\bar{C}_D$, r.m.s.\ lift coefficient $C_L'$, Strouhal number $\mathit{St}$, and mean separation angle $\bar{\theta}_{\mathrm{sep}}$. The results are summarised in table~\ref{tab:results}.

\begin{table}
  \begin{center}
  \def~{\hphantom{0}}
  \setlength{\tabcolsep}{15pt}
  \begin{tabular}{lcccc}
    \toprule
    Case & $\bar{C}_D$ & $C_L'$ & $\mathit{St}$ & $\bar{\theta}_{\mathrm{sep}}$ (deg.) \\
    \midrule
    Standard no-slip        & 1.4119 & 0.2959 & 0.1915 & 66.3 \\
    2nd-order extrapolation & 1.5618 & 0.6856 & 0.2229 & 75.0 \\
    3rd-order extrapolation & 1.2244 & 0.3869 & 0.2279 & 71.9 \\
    \midrule
    \citet{wieselsberger1922}              & 1.3700     & --     & --     & -- \\
    \citet{norberg2003fluctuating}              & --     & 0.3180     & 0.2010     & -- \\
    \citet{gresho1984modified} ($\Rey = 200$) & --     & 0.7400     & --     & -- \\
    \citet{williamson1996vortex}              & --     & --     & 0.1900     & -- \\
    \citet{grove1964experimental}              & --     & --     & --     & 67.0 \\
    \bottomrule
  \end{tabular}
  \caption{Summary of key flow quantities for a circular cylinder at $\Rey = 170$ with imposed no-slip and cases employing varying orders of velocity extrapolation to the wall, compared against the literature. $\bar{C}_D$ is the mean drag coefficient, $C_L'$ is the root-mean-square lift coefficient, $\mathit{St}$ is the Strouhal number, and $\bar{\theta}_{\mathrm{sep}}$ is the mean separation angle. All literature values are quoted at $\Rey = 170$, except where noted.}
  \label{tab:results}
  \end{center}
\end{table}

The no-slip simulation yields good agreement with experimental data for the Strouhal number and mean separation angle. The mean drag coefficient is overestimated relative to experiment, which is consistent with the first-order temporal accuracy of the method; both the explicit Euler integration and Chorin's pressure--velocity splitting incur errors of $O(\Delta t)$ (\citet{Projection_Review}) to which the integrated, pressure-based loads are more sensitive than the frequency and geometric measures of the flow. Regarding the r.m.s.\ lift coefficient, we note that the literature reports a considerable spread in values (see the compilation in \citet{norberg2003fluctuating}), which makes definitive single-reference validation inherently difficult.

\begin{figure}
    \centering
    \includegraphics[width=1.0\textwidth]{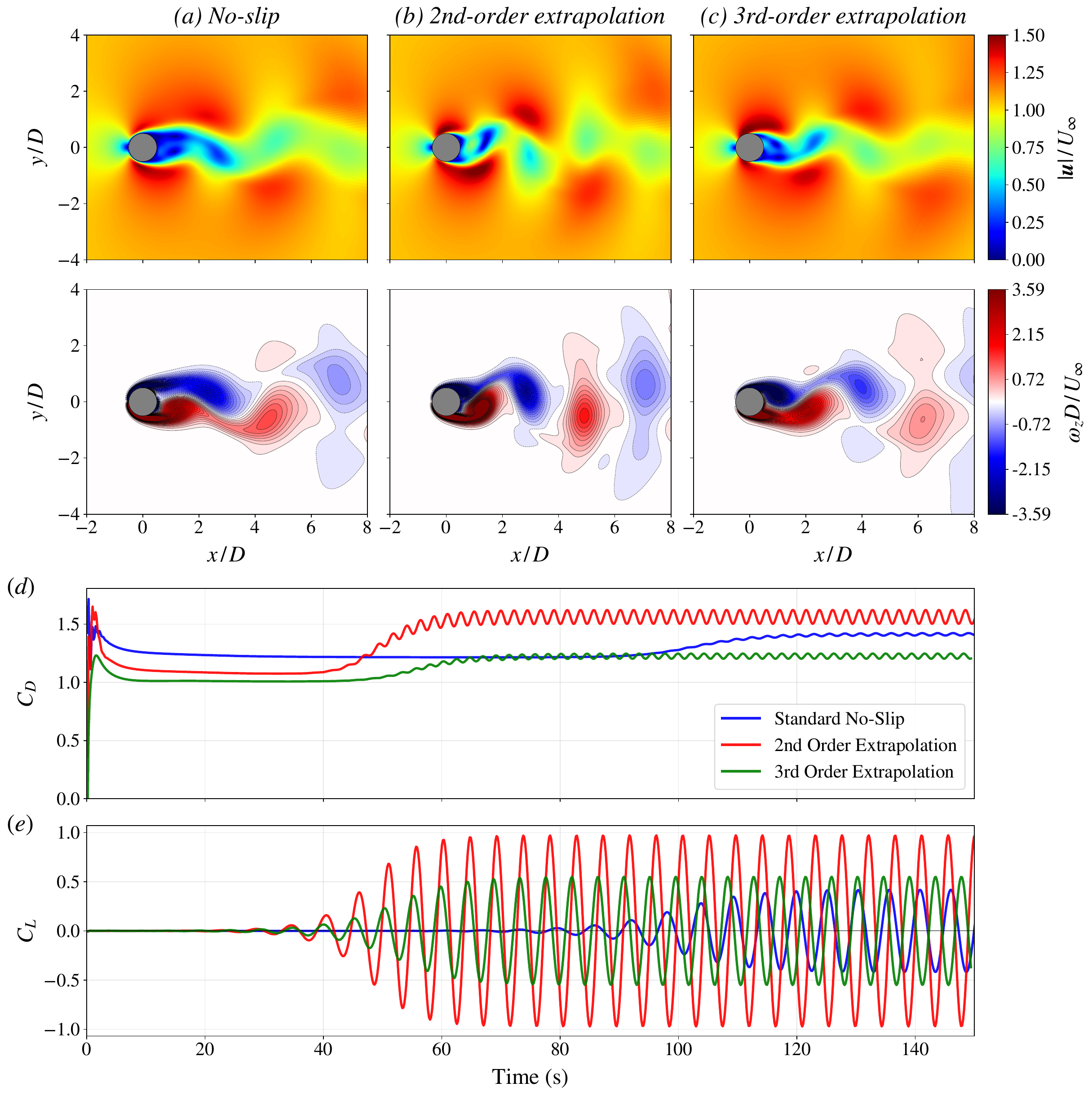}
    \caption{Instantaneous normalised velocity magnitude $|\boldsymbol{u}|\,/\,U_\infty$ (upper panels) and non-dimensional vorticity $\omega_z D\,/\,U_\infty$ (lower panels) for (\textit{a}) the standard no-slip, (\textit{b}) second-order extrapolation, and (\textit{c}) third-order extrapolation cases at $\Rey = 170$, together with the time histories of (\textit{d}) the drag coefficient $C_D$ and (\textit{e}) the lift coefficient $C_L$.}
    \label{fig:force_history}
\end{figure}

Figure~\ref{fig:force_history} presents a comparison between the standard simulation imposing the no-slip boundary condition and the non-sticky-wall simulations using second- and third-order extrapolation, in terms of time histories of the lift and drag coefficients, as well as snapshots of the velocity and vorticity fields. 

The figure shows that the non-sticky-wall simulations lead to vorticity generation and periodic shedding that are qualitatively similar to the no-slip results. Quantitatively, however, the integral flow quantities differ from those of the standard no-slip simulation, with the Strouhal number overestimated in both extrapolation cases, as shown in table~\ref{tab:results}. 

\begin{figure}
    \centering
    \includegraphics[width=0.96\textwidth]{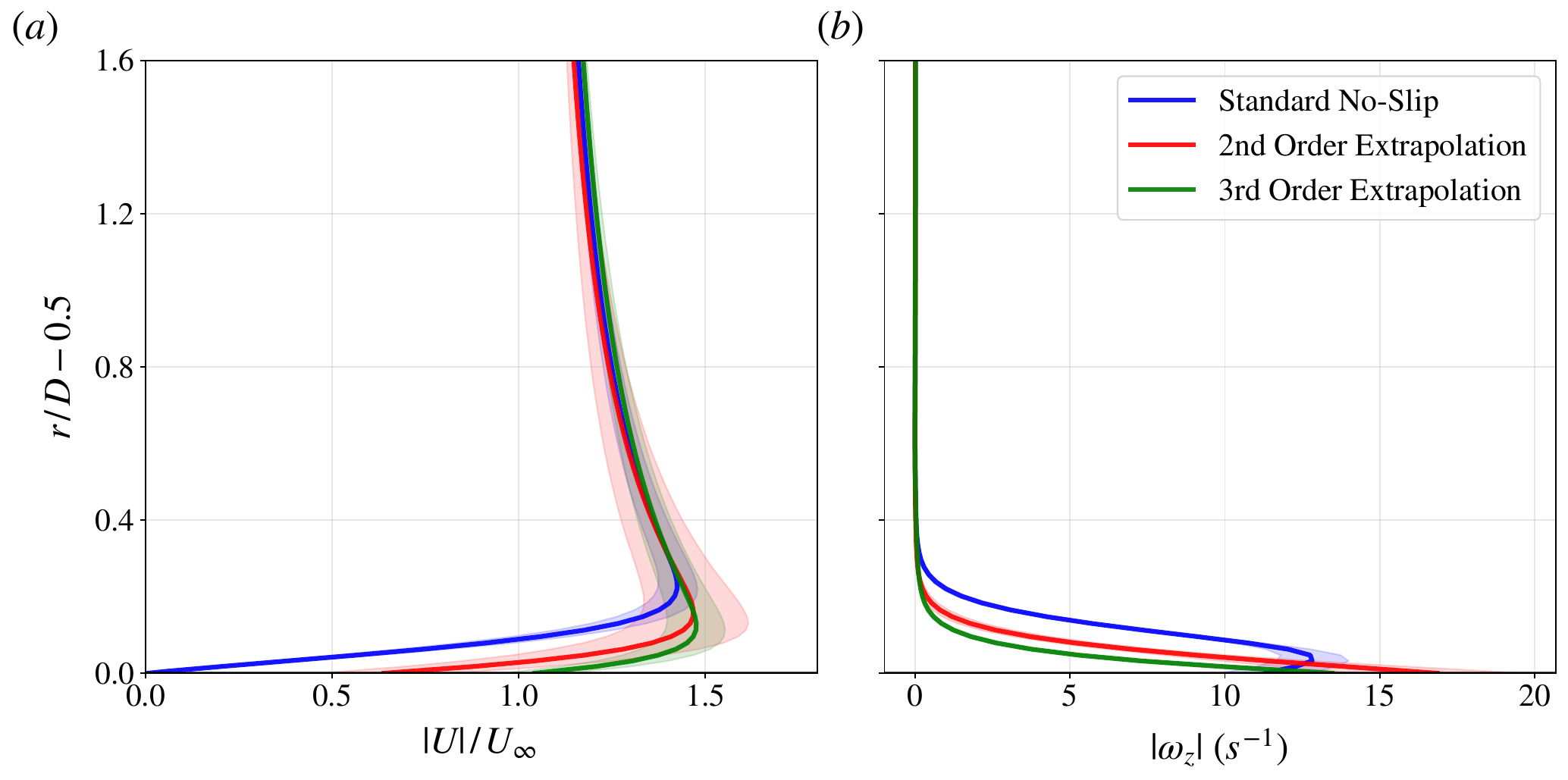}
    \caption{Comparison of time-averaged (\textit{a}) velocity and (\textit{b}) vorticity profiles at the top of the cylinder
    ($\theta = 90^{\circ}$) over one vortex shedding cycle between the standard no-slip simulation and cases employing varying orders of velocity extrapolation to the wall. The shaded regions indicate the temporal variation of each respective quantity over one shedding cycle.}
    \label{fig:velocity_vorticity_profiles}
\end{figure}

It is important to investigate the extent to which slip occurs at the wall in the non-sticky-wall simulations. Figure~\ref{fig:velocity_vorticity_profiles} shows the time-averaged radial profiles of velocity and vorticity magnitudes at $\theta = 90^{\circ}$, computed over one full vortex shedding cycle; the shaded regions indicate the temporal variation of each quantity over the cycle. The figure shows clear slip at the wall in non-sticky-wall simulations, with the amount of slip increasing with the extrapolation order.

We then investigate the extent to which the outer flow (outside the boundary layer) is altered by the wall treatment. The boundary-layer thickness at a given station is defined as the radial distance from the wall at which the vorticity magnitude drops to 1\% of its maximum value; at $\theta=90^\circ$ this yields $0.34\,D$ for the standard no-slip case, and $0.26\,D$ for both the second- and third-order extrapolation cases. The velocity distributions are then sampled along a circular arc concentric with the cylinder and tangent to the boundary-layer edge at $\theta = 90^{\circ}$. Although this circle does not coincide with the boundary-layer edge at every azimuthal station, it remains just outside the boundary layer and therefore suffices for the present purpose.

\begin{figure}
    \centering
    \includegraphics[width=1.0\textwidth]{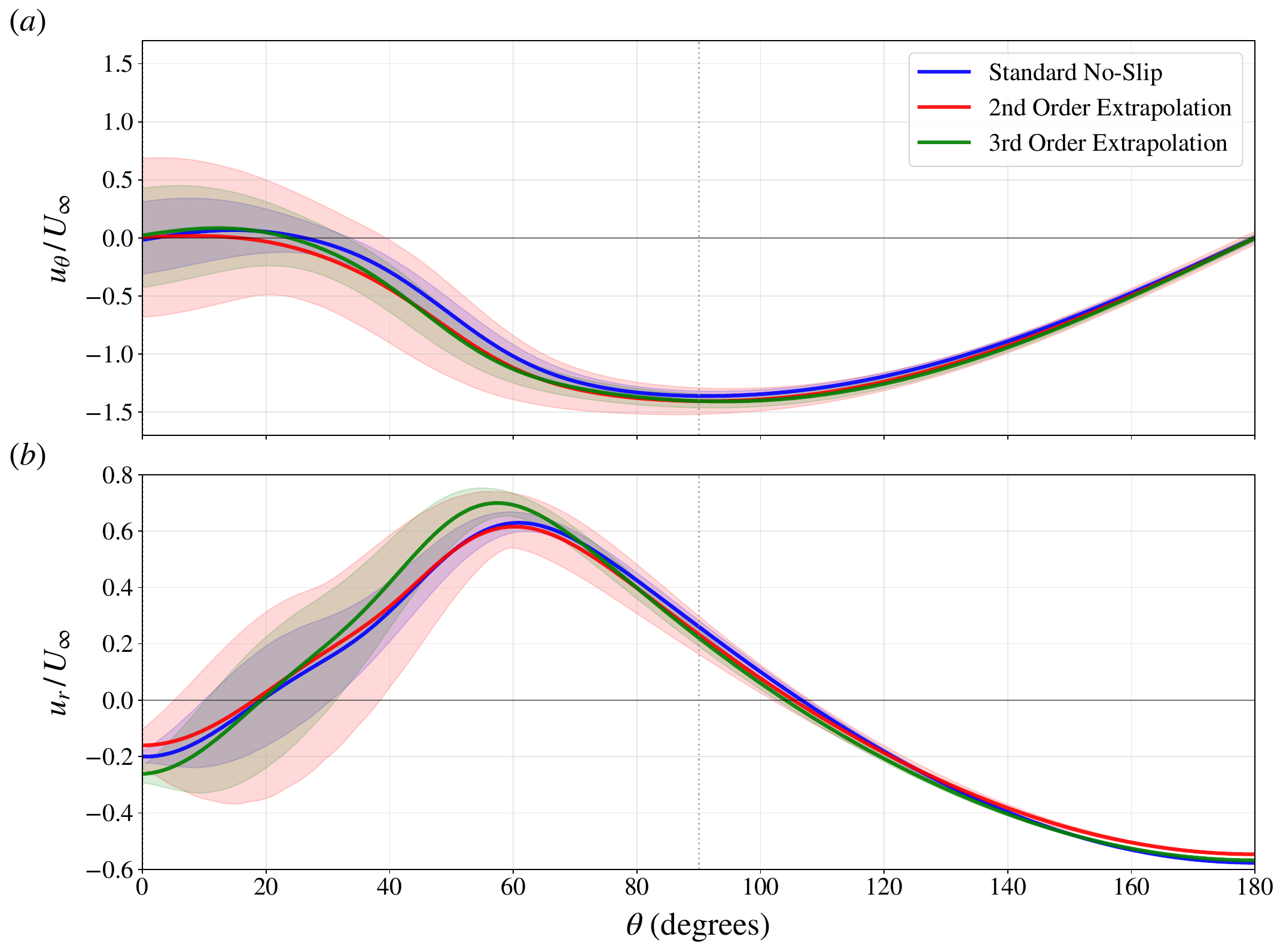}
    \caption{Comparison of time-averaged (\textit{a}) tangential and (\textit{b}) radial velocity distributions, sampled just outside the boundary layer over one vortex shedding cycle, between the standard no-slip simulation and cases employing varying orders of velocity extrapolation to the wall. The shaded region indicates the temporal variation of the velocity magnitude during the cycle.}
    \label{fig:velocity_distributions}
\end{figure}

Figure~\ref{fig:velocity_distributions} shows the time-averaged velocity components sampled along this circle over one full shedding cycle. The outer velocity distributions are in close agreement across all three cases, indicating that although the physics of the boundary layer is not captured by the non-sticky-wall treatment, the flow field outside the boundary layer can be reproduced with reasonable accuracy. This result is particularly important for the present investigation, since, for a stationary body, the edge velocity $U_e$ dictates the total amount of vorticity generated according to the Lighthill mechanism:
\begin{equation}
   \Gamma = \int \int_0^\delta \omega(x,y) dy dx = \int \gamma(x)dx = \int U_e(x) dx. 
 \label{eq:total_vorticity}
\end{equation}
The edge velocity, integrated over half of the cylinder ($0\leq\theta\leq\pi$) and normalised by $U_\infty D$, yields a circulation of 1.90, 1.87, and 1.85 for the no-slip, second-order, and third-order extrapolation simulations, respectively (i.e., less than 3\% deviation). Thus, the total amount of vorticity generated according to the Lighthill mechanism (\ref{eq:total_vorticity}) is found to be remarkably insensitive to the tangential wall treatment, consistent with the assertions of \citet{morton1984generation} and \citet{terrington2020generation}.

We finally examine the separation angle for all three cases. Typically, the separation angle is characterised by the zero-crossing of the skin-friction coefficient along the cylinder wall. However, this criterion is not applicable to the non-sticky-wall simulations, for which the wall shear stress is not meaningful in the conventional no-slip sense. Instead, the separation angle is determined by particle tracing, following an approach commonly employed in experimental studies (\citet{grove1964experimental}).

Particles are released at 60 evenly spaced time instants spanning four shedding cycles, with two particles released simultaneously at each instant from the vicinity of the front stagnation point at $\theta = 180^\circ \pm \varepsilon$. Their trajectories are subsequently tracked, and separation is declared when a particle's radial distance from the cylinder wall exceeds a prescribed threshold. This threshold is calibrated such that the average separation angle obtained from particle tracing in the no-slip case matches that determined from the time-averaged skin-friction coefficient. The same calibrated threshold is then applied to the non-sticky-wall simulations to determine their respective separation angles; this threshold corresponds to a radial distance of approximately 2.5\% of the cylinder diameter from the wall. Sample particle trajectories are shown in the supplementary materials (figure S2).

\begin{figure}
    \centering
    \includegraphics[width=1.0\textwidth]{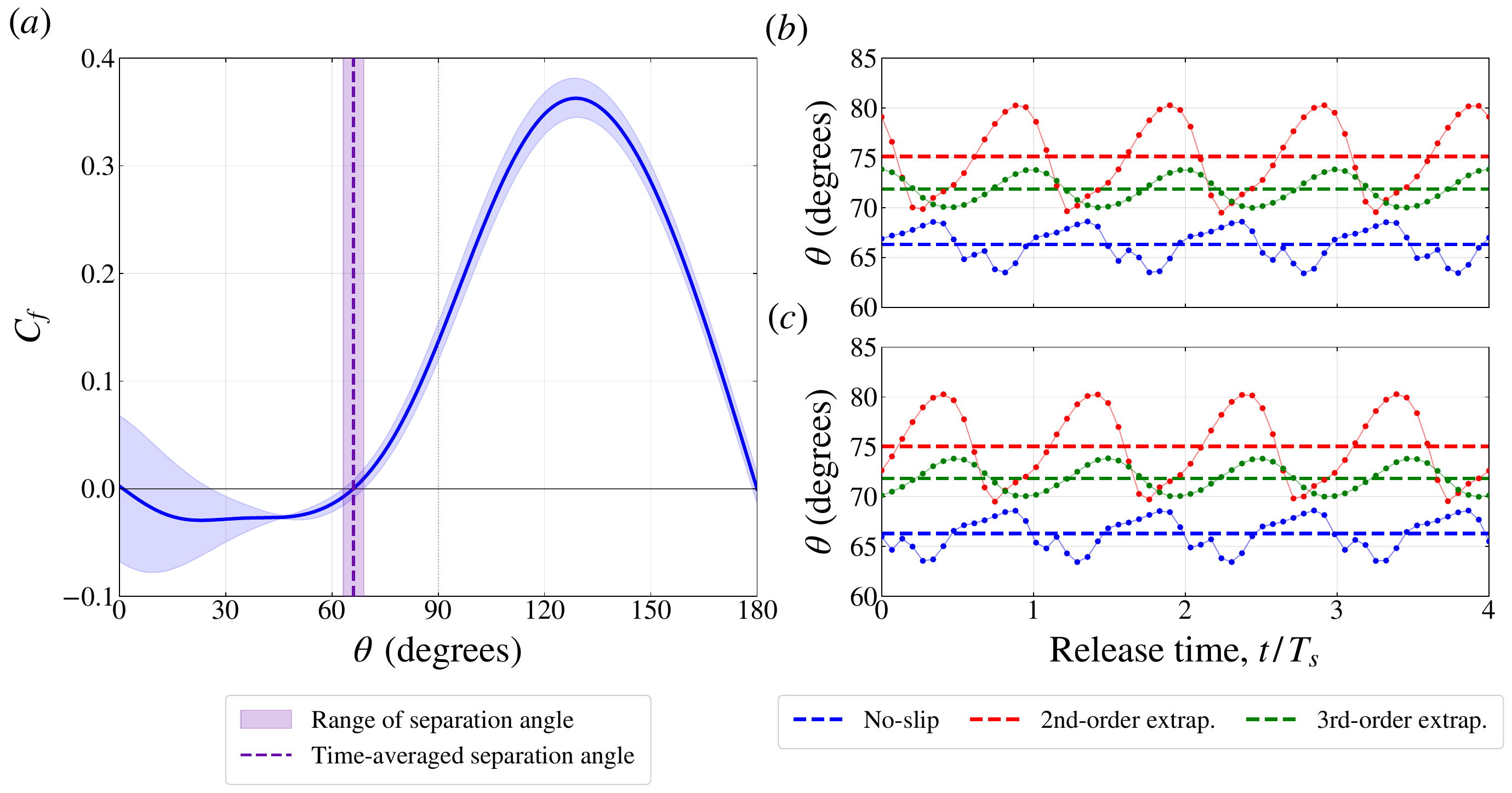}
    \caption{(\textit{a}) Time-averaged skin-friction coefficient $C_f$ distribution along the cylinder wall for the no-slip case; the shaded band indicates the variation over one shedding cycle, and the dashed vertical line with its shaded region denotes the time-averaged separation angle and its cyclic range. (\textit{b},\textit{c}) Instantaneous separation angle as a function of non-dimensional particle release time $t/T_s$ (where $T_s$ is the shedding period), determined by particle tracing for particles released above ($\theta = 180^\circ - \varepsilon$) and below ($\theta = 180^\circ + \varepsilon$) the axis of symmetry, respectively; dashed horizontal lines indicate the time-averaged separation angle for each case (values in table~\ref{tab:results}).}
    \label{fig:separation_angle}
\end{figure}

The results are presented in figure~\ref{fig:separation_angle}. Notably, separation is observed in the non-sticky-wall simulations, even though a classical no-slip boundary layer is absent. However, the separation angles in the non-sticky-wall simulations are higher than that of the no-slip simulation, with higher-order extrapolation yielding closer agreement with the no-slip result.

The above results imply that the no-slip condition is not \textbf{necessary} for vorticity generation and periodic shedding; i.e., it is not the only enabling mechanism, consistent with the assertions of \citet{morton1984generation} and \citet{terrington2020generation}. When the no-slip condition is relaxed, the same qualitative behaviour persists. However, for accurate quantitative prediction of flow quantities in the $\Rey$ regime considered here, the no-slip condition may still be necessary. Whether such necessity for quantitative accuracy persists at high $\Rey$ requires further investigation.

It should be noted that closely related observations were made by \citet{hoffman2010resolution}. In their thought-provoking \textit{Resolution of d'Alembert's Paradox}, the authors found ``\textit{by computational solution of the Euler equations [with added dissipation] with slip boundary conditions, that the potential solution develops into a turbulent solution with substantial drag/lift .... We thus obtain without any presence of viscous boundary layers, substantial drag/lift in accordance with observations and the conjecture of Birkhoff}.'' Similarly, \citet[Sec.~19.4]{Hirsch_Book} wrote: ``\textit{It appeared that computations based on the Euler equations [with artificial dissipation], and which did not implement any form of Kutta condition, still produced accurate results ... of circulation. This has been confirmed since then by many computations on two-dimensional as well as three-dimensional airfoils and wings; see for instance, \citet{Inviscid_Lift_CFD1,Inviscid_Lift_CFD2}.}''

Although these simulations employed some sort of artificial dissipation, they did not impose the no-slip boundary condition. In fact, aside from the use of artificial dissipation, their simulations could be considered more faithful to the numerical experiment under investigation here, since they did not impose any tangential boundary closure whatsoever. Their dissipation terms did not take the form of simple viscous diffusion and, hence, did not require a corresponding tangential wall closure. Without the no-slip boundary condition, or any alternative tangential condition at the wall, their solutions nevertheless developed vorticity. This generation has usually been attributed to artificial dissipation (\citet{Hirsch_Book,Rizzi_Book}). We shall not comment on this attribution, since it is not the main focus of the present paper; our simulations retain the full viscous diffusion term. Rather, we emphasise that these earlier simulations did not implement any form of no-slip or alternative tangential wall closure, yet vorticity was generated. Hence, the no-slip boundary condition cannot by itself be regarded as a \textit{necessary} enabler of vorticity generation, as implied in the classical expositions of \citet{lighthill1963introduction,batchelor1967introduction} and stated more explicitly by \citet{Wu_Vorticity_Generation,De3z_Seeny_Vorticity_Generation,De3z_Seeny_Kutta}.

\section{Conclusions}
\label{sec:conclusions}
The goal of this paper was to investigate whether the no-slip boundary condition (i.e., stickiness) is necessary for vorticity generation, as might be implied by the Lighthill vorticity generation mechanism, or whether vorticity generation is indeed independent of the tangential condition at the wall, as asserted by \citet{morton1984generation} and \citet{terrington2020generation}. To achieve this goal, we investigated the role of the no-slip boundary condition in laminar periodic vortex shedding over a circular cylinder at $\Rey=170$ by replacing the prescribed wall tangential velocity with a dynamically evolving value obtained through second- and third-order extrapolation from the interior flow field. Equivalently, the normal derivative of the tangential velocity at the wall is closed using a one-sided finite-difference approximation in terms of the velocity at nearby interior points. Although this closure may be viewed as imposing \textit{some} tangential boundary condition, it imposes no kinematic stickiness at the wall. The tangential wall velocity is not prescribed, but rather evolves freely according to the flow dynamics, without being constrained to adhere to the surface.

Both the no-slip and non-sticky-wall simulations showed qualitatively similar vorticity dynamics, producing the classical von K\'{a}rm\'{a}n vortex street. This result shows that the no-slip condition is not a prerequisite for vorticity generation or periodic shedding in this flow. The time-averaged velocity distributions just outside the boundary layer are in close agreement across all three cases, indicating that the outer flow is captured with reasonable fidelity even when the wall tangential velocity is not prescribed. This result is particularly important for the present study, since the edge-velocity distribution determines the circulation along a contour at the edge of the boundary layer, which dictates the total amount of vorticity generated at the wall in the no-slip case according to the Lighthill mechanism. This circulation is found to be remarkably insensitive to the wall closure, with deviations of less than 3\% among the three simulations.

Quantitatively, however, the extrapolation cases exhibit non-negligible discrepancies in the mean drag coefficient, r.m.s.\ lift coefficient, and Strouhal number relative to both the no-slip simulation and the literature. In particular, the Strouhal number is overestimated in both extrapolation cases, indicating that the altered near-wall vorticity distribution, even when the outer flow is reproduced with reasonable fidelity, modifies the shedding dynamics. 

The current findings support the argument of \citet{morton1984generation} and \citet{terrington2020generation} that the no-slip condition is not a necessary enabling mechanism for vorticity generation and shedding, although removing it degrades the accuracy of the integrated force and shedding-frequency predictions under the present extrapolation closures. A natural next step could therefore be to replace the extrapolation closure with a first-principles criterion, such as the principle of minimum pressure gradient (\citet{taha2023minimization}), through which the wall slip may be selected dynamically rather than prescribed \textit{a priori}.



\begin{bmhead}[Funding.]
The authors acknowledge support from the National Science Foundation under Grant No. CBET-2332556.
\end{bmhead}

\begin{bmhead}[Declaration of interests.]
The authors report no conflict of interest.
\end{bmhead}

\begin{bmhead}[Data availability statement.]
The data that support the findings of this study are available upon reasonable request.
\end{bmhead}

\begin{bmhead}[Author ORCIDs.]

Haithem E. Taha, https://orcid.org/0000-0002-9242-2045;

Kourosh J. Ghalejooghi, https://orcid.org/0009-0006-5500-1889.
\end{bmhead}


\bibliographystyle{jfm}
\bibliography{references}

\begin{thebibliography}{32}
\expandafter\ifx\csname natexlab\endcsname\relax\def\natexlab#1{#1}\fi
\def\au#1{#1} \def\ed#1{#1} \def\yr#1{#1}\def\at#1{#1}\def\jt#1{\textit{#1}} \def\bt#1{#1}\def\bvol#1{\textbf{#1}} \def\vol#1{#1} \def\pg#1{#1} \def\publ#1{#1}\def\arxiv#1{#1}\def\org#1{#1}\def\st#1{\textit{#1}}

\bibitem[Batchelor(1967)]{batchelor1967introduction}
{\sc \au{Batchelor, George~Keith}} \yr{1967} {\em An introduction to fluid dynamics\/}.  \publ{Cambridge University Press}.

\bibitem[Br{\o}ns {\em et~al.\/}(2014)Br{\o}ns, Thompson, Leweke \& Hourigan]{brons2014vorticity}
{\sc \au{Br{\o}ns, Morten}, \au{Thompson, Mark~Christopher}, \au{Leweke, Thomas} \& \au{Hourigan, Kerry}} \yr{2014}  \at{Vorticity generation and conservation for two-dimensional interfaces and boundaries}.  \jt{Journal of Fluid Mechanics}  \bvol{758},  \pg{63--93}.

\bibitem[Chen {\em et~al.\/}(2024)Chen, Wang \& Liu]{De3z_Seeny_Vorticity_Generation}
{\sc \au{Chen, T.}, \au{Wang, C.} \& \au{Liu, T.}} \yr{2024}  \at{On physics of boundary vorticity creation in incompressible viscous flow}.  \jt{Acta Mechanica Sinica}  \bvol{40}~(3),  \pg{323443}.

\bibitem[Chorin(1968)]{chorin1968numerical}
{\sc \au{Chorin, Alexandre~Joel}} \yr{1968}  \at{Numerical solution of the {Navier--Stokes} equations}.  \jt{Mathematics of Computation}  \bvol{22}~(104),  \pg{745--762}.

\bibitem[Goldstein(1938)]{goldstein1938modern}
{\sc \au{Goldstein, Sydney}} \yr{1938} {\em Modern developments in fluid dynamics: an account of theory and experiment relating to boundary layers, turbulent motion and wakes, vol.~2\/}.  \publ{Clarendon Press}.

\bibitem[Gresho {\em et~al.\/}(1984)Gresho, Chan, Lee \& Upson]{gresho1984modified}
{\sc \au{Gresho, P.~M.}, \au{Chan, S.~T.}, \au{Lee, R.~L.} \& \au{Upson, C.~D.}} \yr{1984}  \at{A modified finite element method for solving the time-dependent, incompressible {N}avier--{S}tokes equations. {P}art 2: applications}.  \jt{International Journal for Numerical Methods in Fluids}  \bvol{4},  \pg{619--640}.

\bibitem[Gresho \& Sani(1987)]{gresho1987pressure}
{\sc \au{Gresho, Philip~M} \& \au{Sani, Robert~L}} \yr{1987}  \at{On pressure boundary conditions for the incompressible {Navier--Stokes} equations}.  \jt{International Journal for Numerical Methods in Fluids}  \bvol{7}~(10),  \pg{1111--1145}.

\bibitem[Grove {\em et~al.\/}(1964)Grove, Shair, Petersen \& Acrivos]{grove1964experimental}
{\sc \au{Grove, A.~S.}, \au{Shair, F.~H.}, \au{Petersen, E.~E.} \& \au{Acrivos, A.}} \yr{1964}  \at{An experimental investigation of the steady separated flow past a circular cylinder}.  \jt{Journal of Fluid Mechanics}  \bvol{19}~(1),  \pg{60--80}.

\bibitem[Guermond {\em et~al.\/}(2006)Guermond, Minev \& Shen]{Projection_Review}
{\sc \au{Guermond, J.-L.}, \au{Minev, P.} \& \au{Shen, J.}} \yr{2006}  \at{An overview of projection methods for incompressible flows}.  \jt{Computer methods in applied mechanics and engineering}  \bvol{195}~(44-47),  \pg{6011--6045}.

\bibitem[Hirsch(1997)]{Hirsch_Book}
{\sc \au{Hirsch, Charles}} \yr{1997} {\em Numerical computation of internal and external flows: computational methods for inviscid and viscous flows, vol.~2\/}.  \publ{Wiley}.

\bibitem[Hirschel {\em et~al.\/}(2021)Hirschel, Rizzi, Breitsamter \& Staudacher]{Rizzi_Book}
{\sc \au{Hirschel, E.~H.}, \au{Rizzi, A.}, \au{Breitsamter, C.} \& \au{Staudacher, W.}} \yr{2021} {\em Separated and Vortical Flow in Aircraft Wing Aerodynamics\/}.  \publ{Springer Berlin Heidelberg}.

\bibitem[Hoffman \& Johnson(2010)]{hoffman2010resolution}
{\sc \au{Hoffman, Johan} \& \au{Johnson, Claes}} \yr{2010}  \at{Resolution of {d'Alembert's} paradox}.  \jt{Journal of Mathematical Fluid Mechanics}  \bvol{12}~(3),  \pg{321--334}.

\bibitem[Lighthill(1963)]{lighthill1963introduction}
{\sc \au{Lighthill, M.~J.}} \yr{1963}  \at{Introduction. {B}oundary layer theory}.  \bt{In {\em Laminar Boundary Layers\/} (ed. \ed{L.~Rosenhead})},  \pg{pp. 46--113}.  \publ{Oxford University Press}.

\bibitem[Mill(1843)]{mill1843system}
{\sc \au{Mill, John~Stuart}} \yr{1843} {\em A System of Logic, Ratiocinative and Inductive\/}.  \publ{London: John W. Parker}.

\bibitem[Morton(1984)]{morton1984generation}
{\sc \au{Morton, Bruce~R}} \yr{1984}  \at{The generation and decay of vorticity}.  \jt{Geophysical \& Astrophysical Fluid Dynamics}  \bvol{28}~(3-4),  \pg{277--308}.

\bibitem[Navier(1823)]{navier1823memoire}
{\sc \au{Navier, C. L. M.~H.}} \yr{1823}  \at{M{\'e}moire sur les lois du mouvement des fluides}.  \jt{M{\'e}moires de l'Acad{\'e}mie Royale des Sciences de l'Institut de France}  \bvol{6},  \pg{389--440}.

\bibitem[Neto {\em et~al.\/}(2005)Neto, Evans, Bonaccurso, Butt \& Craig]{neto2005boundary}
{\sc \au{Neto, Chiara}, \au{Evans, Drew~R}, \au{Bonaccurso, Elmar}, \au{Butt, Hans-J{\"u}rgen} \& \au{Craig, Vincent~SJ}} \yr{2005}  \at{Boundary slip in {N}ewtonian liquids: a review of experimental studies}.  \jt{Reports on Progress in Physics}  \bvol{68}~(12),  \pg{2859--2897}.

\bibitem[Norberg(2003)]{norberg2003fluctuating}
{\sc \au{Norberg, Christoffer}} \yr{2003}  \at{Fluctuating lift on a circular cylinder: review and new measurements}.  \jt{Journal of Fluids and Structures}  \bvol{17}~(1),  \pg{57--96}.

\bibitem[Rajani {\em et~al.\/}(2009)Rajani, Kandasamy \& Majumdar]{rajani2009numerical}
{\sc \au{Rajani, BN}, \au{Kandasamy, A} \& \au{Majumdar, Sekhar}} \yr{2009}  \at{Numerical simulation of laminar flow past a circular cylinder}.  \jt{Applied Mathematical Modelling}  \bvol{33}~(3),  \pg{1228--1247}.

\bibitem[Rizzi(1982)]{Inviscid_Lift_CFD1}
{\sc \au{Rizzi, A.}} \yr{1982}  \at{Damped euler-equation method to compute transonic flow around wing-body combinations}.  \jt{AIAA Journal}  \bvol{20}~(10),  \pg{1321--1328}.

\bibitem[Salazar \& Liu(2025)]{De3z_Seeny_Kutta}
{\sc \au{Salazar, D.~M.} \& \au{Liu, T.}} \yr{2025}  \at{Birth of starting vortex and establishment of {K}utta condition}.  \jt{European Journal of Mechanics-B/Fluids}  \bvol{110},  \pg{19--24}.

\bibitem[Stokes(1845)]{stokes1845theories}
{\sc \au{Stokes, G.~G.}} \yr{1845}  \at{On the theories of the internal friction of fluids in motion, and of the equilibrium and motion of elastic solids}.  \jt{Transactions of the Cambridge Philosophical Society}  \bvol{8},  \pg{287--319}.

\bibitem[Stokes(1851)]{stokes1851effect}
{\sc \au{Stokes, G.~G.}} \yr{1851}  \at{On the effect of the internal friction of fluids on the motion of pendulums}.  \jt{Transactions of the Cambridge Philosophical Society}  \bvol{9},  \pg{8--106}.

\bibitem[Taha {\em et~al.\/}(2023)Taha, Gonzalez \& Shorbagy]{taha2023minimization}
{\sc \au{Taha, Haithem}, \au{Gonzalez, Cody} \& \au{Shorbagy, Mohamed}} \yr{2023}  \at{A minimization principle for incompressible fluid mechanics}.  \jt{Physics of Fluids}  \bvol{35}~(12).

\bibitem[Taha \& Rezaei(2019)]{Viscous_Freq_Resp}
{\sc \au{Taha, H.} \& \au{Rezaei, A.~S.}} \yr{2019}  \at{Viscous extension of potential-flow unsteady aerodynamics: The lift frequency response problem}.  \jt{Journal of Fluid Mechanics}  \bvol{868},  \pg{141--175}.

\bibitem[Terrington {\em et~al.\/}(2020)Terrington, Hourigan \& Thompson]{terrington2020generation}
{\sc \au{Terrington, SJ}, \au{Hourigan, K} \& \au{Thompson, MC}} \yr{2020}  \at{The generation and conservation of vorticity: deforming interfaces and boundaries in two-dimensional flows}.  \jt{Journal of Fluid Mechanics}  \bvol{890},  \pg{A5}.

\bibitem[Wieselsberger(1922)]{wieselsberger1922}
{\sc \au{Wieselsberger, C.}} \yr{1922}  \bt{New data on the laws of fluid resistance}. {\em Tech. Rep.\/} NACA-TN-84.  \org{National Advisory Committee for Aeronautics}.

\bibitem[Williamson(1996)]{williamson1996vortex}
{\sc \au{Williamson, C. H.~K.}} \yr{1996}  \at{Vortex dynamics in the cylinder wake}.  \jt{Annual Review of Fluid Mechanics}  \bvol{28},  \pg{477--539}.

\bibitem[Wu \& Wu(1998)]{Wu_Vorticity_Generation}
{\sc \au{Wu, J.~Z.} \& \au{Wu, J.~M.}} \yr{1998}  \at{Boundary vorticity dynamics since {L}ighthill's 1963 article: review and development}.  \jt{Theoretical and computational fluid dynamics}  \bvol{10}~(1),  \pg{459--474}.

\bibitem[Yoshihara {\em et~al.\/}(1985)Yoshihara, Norstrud, Boerstoel, Chiocchia \& Jones]{Inviscid_Lift_CFD2}
{\sc \au{Yoshihara, H.}, \au{Norstrud, H.}, \au{Boerstoel, J.~W.}, \au{Chiocchia, G.} \& \au{Jones, D.~J.}} \yr{1985}  \bt{Test cases for inviscid flow field methods}. {\em Tech. Rep.\/} AR-211.  \org{AGARD}.

\bibitem[Young(1954)]{young1954iterative}
{\sc \au{Young, David}} \yr{1954}  \at{Iterative methods for solving partial difference equations of elliptic type}.  \jt{Transactions of the American Mathematical Society}  \bvol{76}~(1),  \pg{92--111}.

\bibitem[Zdravkovich(1997)]{zdravkovich1997flow}
{\sc \au{Zdravkovich, Momchilo~M}} \yr{1997} {\em Flow around circular cylinders: a comprehensive guide through flow phenomena, experiments, applications, mathematical models, and computer simulations\/}.  \publ{Oxford University Press}.

\end{thebibliography}
\end{document}